\documentclass[fleqn,11pt,twoside]{article}

\usepackage{amsthm,amsthm,amssymb, color, xcolor,epsfig, graphics, subfigure}

\usepackage{amsmath, graphicx, latexsym, lscape }

\makeatletter
\newcommand{\copyrightnote}[2]{{\renewcommand{\thefootnote}{}
 \footnotetext{\small\it
\begin{flushleft}
 \copyright \ #1   #2  
\end{flushleft}}}}

\newcommand{\Name}[1]{\begin{flushleft}
                       \LARGE \bf #1
                       \end{flushleft}\vspace{-3mm}}

\newcommand{\Author}[1]{\begin{flushleft}
                       \it #1 \end{flushleft}}

\newcommand{\Address}[1]{\begin{flushleft}
                       \it #1 \end{flushleft}}

\newcommand{\Date}[1]{\begin{flushleft}
                      \small  \it #1 \end{flushleft}}

\newcommand{\evenhead}{Author \ name}
\newcommand{\oddhead}{Article \ name}

\renewcommand{\@evenhead}{
\hspace*{-3pt}\raisebox{-15pt}[\headheight][0pt]{\vbox{\hbox to \textwidth
{\thepage \hfil \evenhead}\vskip4pt \hrule}}}
\renewcommand{\@oddhead}{
\hspace*{-3pt}\raisebox{-15pt}[\headheight][0pt]{\vbox{\hbox to \textwidth
{\oddhead \hfil \thepage}\vskip4pt\hrule}}}
\renewcommand{\@evenfoot}{}
\renewcommand{\@oddfoot}{}

\long\def\@makecaption#1#2{%
  \vskip\abovecaptionskip
  \sbox\@tempboxa{\small \textbf{#1.}\ \ #2}%
  \ifdim \wd\@tempboxa >\hsize
    {\small \textbf{#1.}\ \ #2}\par
  \else
    \global \@minipagefalse
    \hb@xt@\hsize{\hfil\box\@tempboxa\hfil}%
  \fi
  \vskip\belowcaptionskip}

\newcommand{\JNMPnumberwithin}[3][\arabic]{%
  \@ifundefined{c@#2}{\@nocounterr{#2}}{%
    \@ifundefined{c@#3}{\@nocnterr{#3}}{%
      \@addtoreset{#2}{#3}%
      \@xp\xdef\csname the#2\endcsname{%
        \@xp\@nx\csname the#3\endcsname .\@nx#1{#2}}}}%
}

\newcommand{\resetfootnoterule} {
  \renewcommand\footnoterule{%
  \kern-3\p@
  \hrule\@width.4\columnwidth
  \kern2.6\p@}
}

\renewcommand{\footnoterule}{}

\makeatother

\theoremstyle{definition}

\begin{document}

\renewcommand{\evenhead}{ {\LARGE\textcolor{blue!10!black!40!green}{{\sf \ \ \ ]ocnmp[}}}\strut\hfill J Yu and Y Feng}
\renewcommand{\oddhead}{ {\LARGE\textcolor{blue!10!black!40!green}{{\sf ]ocnmp[}}}\ \ \ \ \  Lie symmetries,  (2+1)-Dimensional time fractional KP equation}

\thispagestyle{empty}
\newcommand{\FistPageHead}[3]{
\begin{flushleft}
\raisebox{8mm}[0pt][0pt]
{\footnotesize \sf
\parbox{150mm}{{Open Communications in Nonlinear Mathematical Physics}\ \  {\LARGE\textcolor{blue!10!black!40!green}{]ocnmp[}}
\ Vol.4 (2024) pp
#2\hfill {\sc #3}}}\vspace{-13mm}
\end{flushleft}}

\FistPageHead{1}{\pageref{firstpage}--\pageref{lastpage}}{ \ \ Article}

\strut\hfill

\strut\hfill

\copyrightnote{The author(s). Distributed under a Creative Commons Attribution 4.0 International License}

\Name{Lie symmetry analysis of (2+1)-dimensional time fractional Kadomtsev-Petviashvili equation}

\Author{Jicheng Yu${^{1,\ast}}$ and Yuqiang Feng${^{1,2}}$} 

\Address{${^{1}}$ School of Science, Wuhan University of Science and Technology, Wuhan 430081, Hubei, China\\
	${^{2}}$ Hubei Province Key Laboratory of Systems Science in
	Metallurgical Process, Wuhan 430081, Hubei,  China\\
	${^{\ast}}$ yjicheng@126.com}

\Date{Received May 20, 2024; Accepted August 13, 2024}

\setcounter{equation}{0}

\begin{abstract}
\noindent 
In this paper, Lie symmetry analysis method is applied to the (2+1)-dimensional time fractional Kadomtsev-Petviashvili (KP) equation with the mixed derivative of Riemann-Liouville time-fractional derivative and integer-order $x$-derivative. We obtained all the Lie symmetries admitted by the KP equation and used them to reduce the (2+1)-dimensional fractional partial differential equation with Riemann-Liouville fractional derivative to some (1+1)-dimensional fractional partial differential equations with Erd\'{e}lyi-Kober fractional derivative or Riemann-Liouville fractional derivative, thereby getting some exact solutions of the reduced equations. In addition, the new conservation theorem and the generalization of Noether operators are developed to construct the conservation laws for the equation studied.
\end{abstract}

\label{firstpage}


\section{Introduction}
 
The (2+1)-dimensional Kadomtsev-Petviashvili equation, named after B. B. Kadomtsev and V. I. Petviashvili \cite{authour1}, is an important nonlinear partial differential equation in mathematical physics. It is given by 
$$u_{xt}-uu_{xx}-u^2_{x}-u_{xxxx}=u_{yy}, \eqno{(1.1)}  $$
which is one of the few integrable equations in high dimensions and comes from the study of long gravity waves in a single layer or multilayered shallow fluid when the waves propagate predominantly in one direction with a small perturbation in the perpendicular one \cite{authour2}. It also appears in many other fields, such as plasma physics, gas dynamics, etc. Due to the widespread application of Eq. (1.1), it has attracted the interest of many scholars. Especially, Chen \cite{authour3} found the Backlund transformations related to symmetries and integrability of the KP equation in 1975. Cheng and Li \cite{authour4} discussed the constraint of the KP equation and its special solutions. Lou \cite{authour5} derived an explicit and simple constructive formula for the symmetries of the KP equation. Biondini and Pelinovsky \cite{authour6} discussed the two-dimensional localized solution and resonant two-soliton solution and their properties for the KP equation. Ma \cite{authour7} presented a class of lump solutions, rationally localized in all directions in the space, to the KP equation through symbolic computation with Maple. In addition, a large number of research results on the KP equation have emerged in recent years (see \cite{authour8,authour9,authour10,authour11,authour12,authour13,authour14,authour15} and the references therein).

In this paper, we applied Lie symmetry analysis method to study the following (2+1)-dimensional time fractional KP equation:
$$D^{\alpha}_{t}u_{x}-uu_{xx}-u^2_{x}-u_{xxxx}=u_{yy},\ \ 0<\alpha<1, \eqno{(1.2)}  $$
with Riemann-Liouville fractional derivative defined by \cite{authour16}
$$ _{a}D_{t}^\alpha f(t,x) =D_{t}^{n}\ _{a}I_{t}^{n-\alpha} f(t,x)= \left\{
\begin{array}{lll}
	\frac{1}{\Gamma (n-\alpha )}\frac{\partial^n}{\partial t^n}\int_a^t {\frac{f(s,x)}{(t-s)^{ \alpha-n+1 }} }\mathrm{d}s,\ \  & n-1<\alpha<n, n\in\mathbb{N} \\
	\\
	D_{t}^{n}f(t,x),\ \  & \alpha=n\in\mathbb{N}  \\
\end{array}\right.$$
for $t>a$. We denote the operator $_{0}D^{\alpha}_{t}$ as $D^{\alpha}_{t}$ throughout this paper, while $D^{-\alpha}_{t}=I^{\alpha}_{t}$ is Riemann-Liouville fractional integral. Recently, many researchers have used different methods and techniques to study the fractional KP equation (see \cite{authour17,authour18,authour19,authour20,authour21,authour22,authour23,authour24} and references therein). Especially, Borluk et al. \cite{authour22} proved the existence of periodically modulated solitary wave solutions of the fractional KP equation by dimension-breaking bifurcation and discussed the line solitary wave solutions and their transverse (in)stability. They also proved existence of lump solutions for the fractional KP equation via a variational approach, and presented numerically generated lump solutions and observed the cross-sectional symmetry of the solutions numerically in \cite{authour23}.

As a generalization of the classical calculus, fractional calculus can be traced back to the letter written by L'H\^{o}spital to Leibniz in 1695. Since then, it has gradually gained the attention of mathematicians. Especially in recent decades, it has developed rapidly and been successfully applied in many fields of science and technology \cite{authour25,authour26,authour27,authour28}. Therefore, it is very important to find the solution of fractional differential equation. So far, there have been some numerical and analytical methods, such as Adomian decomposition method \cite{authour29}, finite difference method \cite{authour30}, homotopy perturbation method \cite{authour31}, the sub-equation method \cite{authour32}, the variational iteration method \cite{authour33}, Lie symmetry analysis method \cite{authour34}, invariant subspace method \cite{authour35} and so on. Among them, Lie symmetry analysis method has received an increasing attention.

Lie symmetry analysis method was founded by Norwegian mathematician Sophus Lie at the end of the nineteenth century and then further developed by some other mathematicians, such as Ovsiannikov \cite{authour36}, Olver \cite{authour37}, Ibragimov \cite{authour38,authour39,authour40} and so on.  As a modern method among many analytic techniques, Lie symmetry analysis has been extended to fractional differential equations (FDEs) by Gazizov et al. \cite{authour34} in 2007. It was then effectively applied to various models of the FDEs occurring in different areas of applied science (see \cite{authour41,authour42,authour43,authour44,authour45,authour46,authour47,authour48,authour49,authour50}). Lie symmetry analysis method can treat differential equations uniformly regardless of their forms, transforming some solutions of these equations into other forms of solutions by means of continuous point transformations \cite{authour51}. The algorithm of its using for fractional partial differential equations (FPDEs) is as follows:

1. To get the symmetry group admitted by the considered FPDEs by constructing local one-parameter continuous point transformations.

2. In order to achieve dimensionality reduction, the obtained group generators are used to perform similarity reductions on FPDEs.

3. For the reduced FDEs, various analytical and numerical methods are used to find their solutions. For high-dimensional reduced FPDEs, go back to procedure one for iteration.

The aim of this paper is to find all Lie symmetries for Eq. (1.2) by using Lie symmetry analysis method. Subsequently, for each Lie symmetry, we reduce Eq. (1.2) to (1+1)-dimensional time fractional partial differential equations or time fractional ordinary differential equations. Moreover, we obtain some exact solutions for the reduced equations and construct the conservation laws by the new conservation theorem and the generalization of Noether operators.

This paper is organized as follows. In Section 2, Lie symmetry analysis of Eq. (1.2) is presented. In Section 3, the similarity reductions and invariant solutions for Eq. (1.2) are obtained. The conserved vectors for all Lie symmetries admitted by Eq. (1.2) are constructed in Sections 4. The conclusion is given in the last section.

\section{Lie symmetry analysis of Eq. (1.2)} 

 Consider the (2+1)-dimensional time fractional Kadomtsev-Petviashvili equation (1.2), which is assumed to be invariant under the one-parameter ($\epsilon $) Lie group of continuous point transformations, i.e.,
\begin{equation}
	\begin{split}
		&t^{*}=t+\epsilon\tau(t,x,y,u)+o(\epsilon),\ \ \ \ \ \ \ \ x^{*}=x+\epsilon\xi(t,x,y,u)+o(\epsilon), \\
		&y^{*}=y+\epsilon\theta(t,x,y,u)+o(\epsilon),\ \ \ \ \ \ \  u^{*}=u+\epsilon\eta(t,x,y,u)+o(\epsilon), \\
		&D^{\alpha}_{t^{*}}u_{x^*}^{*}=D^{\alpha}_{t}u_{x}+\epsilon\eta^{\alpha,1}+o(\epsilon),\ \ \ \ D_{x^{*}}u^{*}=D_{x}u+\epsilon\eta^{x}+o(\epsilon), \\
		&D_{y^{*}}u^{*}=D_{y}u+\epsilon\eta^{y}+o(\epsilon),\ \ \ \ \ \ \ \ \
		D^{2}_{x^{*}}u^{*}=D^{2}_{x}u+\epsilon\eta^{xx}+o(\epsilon),\\
		&D^{2}_{y^{*}}u^{*}=D^{2}_{y}u+\epsilon\eta^{yy}+o(\epsilon),\ \ \ \ \ \ \ \ D^{4}_{x^{*}}u^{*}=D^{4}_{x}u+\epsilon\eta^{xxxx}+o(\epsilon),
	\end{split} \tag{2.1}
\end{equation}
where $\tau$, $\xi$, $\theta$, $\eta$ are infinitesimals, and $\eta^{\alpha,1}$, $\eta^{x}$, $\eta^{y}$, $\eta^{xx}$, $\eta^{yy}$, $\eta^{xxxx}$ are the corresponding prolongations of $\eta$. The group generator is defined by
$$X=\tau\frac{\partial }{\partial t }+\xi\frac{\partial }{\partial x }+\theta\frac{\partial }{\partial y }+\eta\frac{\partial }{\partial u }, \eqno{(2.2)}$$
and its prolongation has the form
$$prX=X+\eta^{\alpha,1}\frac{\partial }{\partial(D^{\alpha}_{t}u_{x})}+\eta^{x}\frac{\partial }{\partial u_{x}}+\eta^{y}\frac{\partial }{\partial u_{y}}+\eta^{xx}\frac{\partial }{\partial u_{xx}}+\eta^{yy}\frac{\partial }{\partial u_{yy}}+\eta^{xxxx}\frac{\partial }{\partial u_{xxxx}}, \eqno{(2.3)}$$
where
$$\eta^{x}=D_{x}(\eta-\tau u_{t}-\xi u_{x}-\theta u_{y})+\tau u_{xt}+\xi u_{xx}+\theta u_{xy},\eqno{(2.4)}$$
$$\eta^{y}=D_{y}(\eta-\tau u_{t}-\xi u_{x}-\theta u_{y})+\tau u_{yt}+\xi u_{yx}+\theta u_{yy},\eqno{(2.5)}$$
$$\eta^{xx}=D^2_{x}(\eta-\tau u_{t}-\xi u_{x}-\theta u_{y})+\tau u_{xxt}+\xi u_{xxx}+\theta u_{xxy},\eqno{(2.6)}$$
$$\eta^{yy}=D^2_{y}(\eta-\tau u_{t}-\xi u_{x}-\theta u_{y})+\tau u_{yyt}+\xi u_{yyx}+\theta u_{yyy},\eqno{(2.7)}$$
$$\eta^{xxxx}=D^4_{x}(\eta-\tau u_{t}-\xi u_{x}-\theta u_{y})+\tau u_{xxxxt}+\xi u_{xxxxx}+\theta u_{xxxxy},\eqno{(2.8)}$$
and \cite{authour52,authour53,authour54}
\begin{equation}
	\begin{split}
		\eta^{\alpha,1}=&D^{\alpha}_{t}[D_{x}(\eta-\tau u_{t}-\xi u_{x}-\theta u_{y})]+\tau D^{\alpha+1}_{t}(u_{x})+\xi D^{\alpha}_{t}(u_{xx})+\theta D^{\alpha}_{t}(u_{xy})\\
		=&D^{\alpha}_{t}(R)-\sum_{n=1}^{\infty }\binom{\alpha }{n+1}D^{n+1}_{t}(\tau)D^{\alpha-n}_{t}(u_{x})\\
		&-\sum_{n=1}^{\infty }\binom{\alpha }{n}D^{n}_{t}(\xi)D^{\alpha-n}_{t}(u_{xx})-\sum_{n=1}^{\infty }\binom{\alpha }{n}D^{n}_{t}(\theta)D^{\alpha-n}_{t}(u_{xy}),\ \
	\end{split} \tag{2.9}
\end{equation}
with
$$R=\eta_x+u_x\eta_u-u_t(\tau_x+u_x\tau_u)-u_x(\xi_x+u_x\xi_u)-u_y(\theta_x+u_x\theta_u),\eqno{(2.10)}$$
where $D_{t}$, $D_{x}$ and $D_{y}$ are the total derivative with respect to $t$, $x$ and $y$, respectively.

The one-parameter Lie symmetry transformations (2.1) are admitted by Eq. (1.2), if the following invariance criterion holds:
$$prX\big(D^{\alpha}_{t}u_{x}-uu_{xx}-u^2_{x}-u_{xxxx}-u_{yy}\big)|_{(1.2)}=0,
\eqno{(2.11)} $$
which can be rewritten as
$$\big(\eta^{\alpha,1}-\eta^{xxxx}-u\eta^{xx}-\eta^{yy}-2u_x\eta^{x}-u_{xx}\eta\big)|_{(1.2)}=0.\eqno{(2.12)} $$
Putting $\eta^{\alpha,1}$, $\eta^{x}$, $\eta^{y}$, $\eta^{xx}$, $\eta^{yy}$ and $\eta^{xxxx}$ into (2.12) and letting coefficients of various derivatives of $u$ to be zero, we can obtain the infinitesimals as follows:
$$\tau=c_1t,\ \ \xi=\frac{\alpha }{3 }c_1x+c_2,\ \ \theta=\frac{2\alpha }{3 }c_1y+c_3, \ \ \eta=-\frac{2\alpha }{3 }c_1u,\eqno{(2.13)}$$
where $c_1$, $c_2$ and $c_3$ are arbitrary constants. So Eq. (1.2) admits the three-dimension Lie algebra spanned by
$$X_1=t\frac{\partial }{\partial t }+\frac{\alpha }{3 }x\frac{\partial }{\partial x }+\frac{2\alpha }{3 } y\frac{\partial }{\partial y }-\frac{2\alpha }{3 }u\frac{\partial }{\partial u },\ \ X_2=\frac{\partial }{\partial x },\ \ X_3=\frac{\partial }{\partial y }.\eqno{(2.14)} $$

\section{Similarity reductions and invariant solutions of Eq. (1.2)} 

In this section, the aimed equation (1.2) can be reduced to some (1+1)-dimensional fractional partial differential equations with the left-hand Erd\'{e}lyi-Kober fractional derivative and some other solvable fractional differential equations with Riemann-Liouville fractional derivative. 

\bigskip
\noindent{\bf Case 1: $X_1=t\frac{\partial }{\partial t }+\frac{\alpha }{3 }x\frac{\partial }{\partial x }+\frac{2\alpha }{3 } y\frac{\partial }{\partial y }-\frac{2\alpha }{3 }u\frac{\partial }{\partial u }$}
\bigskip

The characteristic equation corresponding to the group generator $X_1$ is
$$\frac{\mathrm{d} t }{t }=\frac{\mathrm{d}x }{\frac{\alpha }{3 }x }=\frac{\mathrm{d} y }{\frac{2\alpha }{3 } y }=\frac{\mathrm{d}u }{-\frac{2\alpha }{3 }u },\eqno{(3.1)}$$
from which, we obtain the similarity variables $xt^{-\frac{\alpha }{3 }}$, $yt^{-\frac{2\alpha }{3 }}$ and $ut^{\frac{2\alpha }{3 }} $. So we get the invariant solution of Eq. (1.2) as follows:
$$u(t,x,y)=t^{-\frac{2\alpha }{3 }}f(\omega_1,\omega_2),\ \ \omega_1=xt^{-\frac{\alpha }{3 }},\ \ \omega_2=yt^{-\frac{2\alpha }{3 }}.\eqno{(3.2)}$$

\noindent{\bf Theorem 3.1. }The similarity transformation $u(t,x,y)=t^{-\frac{2\alpha }{3 }}f(\omega_1,\omega_2)$ with the similarity variables $\omega_1=xt^{-\frac{\alpha }{3 }}$, $\omega_2=yt^{-\frac{2\alpha }{3 }}$ reduce Eq. (1.2) to (1+1)-dimensional fractional partial differential equations given by
$$(\mathcal{P}^{1-2\alpha,\alpha}_{\frac{3 }{\alpha},\frac{3 }{2\alpha}}f_{\omega_1})(\omega_1,\omega_2)-ff_{\omega_1\omega_1}-f^2_{\omega_1}-f_{\omega_1\omega_1\omega_1\omega_1}-f_{\omega_2\omega_2}=0,\eqno{(3.3)} $$
where $(\mathcal{P}^{\iota,\kappa}_{\delta_1,\delta_2})$ is the left-hand Erd\'{e}lyi-Kober fractional differential operator defined by
$$(\mathcal{P}^{\iota,\kappa}_{\delta_1,\delta_2}\psi)(\omega_1,\omega_2):=\prod_{j=0}^{m-1}(\iota+j-\frac{1 }{\delta_1}\omega_1\frac{\mathrm{d} }{\mathrm{d}\omega_1 }-\frac{1 }{\delta_2}\omega_2\frac{\mathrm{d} }{\mathrm{d}\omega_2 })(\mathcal{K}^{\iota+\kappa,m-\kappa}_{\delta_1,\delta_2}\psi)(\omega_1,\omega_2),\ \kappa>0,  $$
$$m=\left\{
\begin{array}{lll}
	[\kappa]+1,\ \ &\mathrm{if} \ \kappa \notin \mathbb{N}, \\
	\kappa,\ \ &\mathrm{if} \ \kappa \in \mathbb{N},  \\
\end{array}\right.\eqno{(3.4)} $$
where
$$(\mathcal{K}^{\iota,\kappa}_{\delta_1,\delta_2}\psi)(\omega_1,\omega_2):=\left\{
\begin{array}{lll}
	\frac{1 }{\Gamma(\kappa) }\int_1^\infty(s-1)^{\kappa-1}s^{-(\iota+\kappa)}\psi(\omega_1 s^{\frac{1 }{\delta_1 }},\omega_2 s^{\frac{1 }{\delta_2 }})\mathrm{d}s,\ \ &\kappa>0, \\
	\psi(\omega_1,\omega_2),\ \ &\kappa=0,  \\
\end{array}\right.\eqno{(3.5)}$$
is the left-hand Erd\'{e}lyi-Kober fractional integral operator.

\noindent{\bf Proof: } For $0<\alpha<1$, the Riemann-Liouville time fractional derivative of $u(t,x,y)$ can be obtained as follows:
$$D^{\alpha}_{t}u_{x}=\frac{\partial^\alpha  }{\partial t^\alpha }(t^{-\alpha}f_{\omega_1}(\omega_1,\omega_2))=\frac{\partial  }{\partial t }\Big[\frac{1}{\Gamma (1-\alpha )}\int_0^t (t-s)^{-\alpha }s^{-\alpha}f_{\omega_1}(xs^{-\frac{\alpha }{3 }},ys^{-\frac{2\alpha }{3 }} )\mathrm{d}s\Big].  $$
Assuming $r=\frac{t }{s}$, we have
$$\frac{\partial  }{\partial t }\Big[\frac{t^{1-2\alpha}}{\Gamma (1-\alpha )}\int_1^\infty (r-1)^{-\alpha }r^{2\alpha-2}f_{\omega_1}(\omega_1 r^{\frac{\alpha }{3 }},\omega_2 r^{\frac{2\alpha }{3 }})\mathrm{d}r\Big]=\frac{\partial  }{\partial t }\Big[t^{1-2\alpha}(\mathcal{K}^{1-\alpha,1-\alpha}_{\frac{3 }{\alpha},\frac{3 }{2\alpha}}f_{\omega_1})(\omega_1,\omega_2)\Big].  $$
Because of $\omega_1=xt^{-\frac{\alpha }{3}}$ and $\omega_2=yt^{-\frac{2\alpha }{3}}$, the following relation holds:
\begin{equation}\nonumber
	t\frac{\partial  }{\partial t }f_{\omega_1}(\omega_1,\omega_2)=-\frac{\alpha }{3}\omega_1\frac{\partial  }{\partial \omega_1 }f_{\omega_1}(\omega_1,\omega_2)-\frac{2\alpha }{3}\omega_2\frac{\partial  }{\partial \omega_2 }f_{\omega_1}(\omega_1,\omega_2).
\end{equation}
Hence, we arrive at
\begin{equation}\nonumber
	\begin{split}
		D^{\alpha}_{t}u_{x}&=t^{-2\alpha}\Big[(1-2\alpha-\frac{\alpha }{3}\omega_1\frac{\partial  }{\partial \omega_1 }-\frac{2\alpha }{3}\omega_2\frac{\partial  }{\partial \omega_2 })(\mathcal{K}^{1-\alpha,1-\alpha}_{\frac{3 }{\alpha},\frac{3 }{2\alpha}}f_{\omega_1})(\omega_1,\omega_2)\Big] \\
		&=t^{-2\alpha}(\mathcal{P}^{1-2\alpha,\alpha}_{\frac{3 }{\alpha},\frac{3 }{2\alpha}}f_{\omega_1})(\omega_1,\omega_2).
	\end{split} 
\end{equation}
Meanwhile,
$$-uu_{xx}-u^2_{x}-u_{xxxx}-u_{yy}=t^{-2\alpha}(-ff_{\omega_1\omega_1}-f^2_{\omega_1}-f_{\omega_1\omega_1\omega_1\omega_1}-f_{\omega_2\omega_2}).$$
This completes the proof. \hfill{$\square$}

Next we use the power series method to derive the power series solution of the reduced equations (3.3). Let us assume 
$$f(\omega_1,\omega_2)=\sum_{n,m=0}^{\infty}a_{n,m} \omega_1^n\omega_2^m,\eqno{(3.6)}$$ 
then
\begin{equation}
	\begin{split}
		&\frac{\partial f }{\partial \omega_1}=\sum_{n,m=0}^{\infty}(n+1)a_{n+1,m} \omega_1^n\omega_2^m, \\
		&\frac{\partial^2 f }{\partial \omega_1^2}=\sum_{n,m=0}^{\infty}(n+2)(n+1)a_{n+2,m}\omega_1^n\omega_2^m,\\ &\frac{\partial^2 f }{\partial \omega_2^2}=\sum_{n,m=0}^{\infty}(m+2)(m+1)a_{n,m+2}\omega_1^n\omega_2^m,\\
		&\frac{\partial^4 f }{\partial \omega_1^4}=\sum_{n,m=0}^{\infty}(n+4)(n+3)(n+2)(n+1)a_{n+4,m}\omega_1^n\omega_2^m,
	\end{split} \tag{3.7}
\end{equation}
and
\begin{equation}
	\begin{split}
		(\mathcal{P}^{1-2\alpha,\alpha}_{\frac{3 }{\alpha},\frac{3 }{2\alpha}}f_{\omega_1})(\omega_1,\omega_2)=&(1-2\alpha-\frac{\alpha }{3}\omega_1\frac{\partial }{\partial \omega_1 }-\frac{2\alpha }{3}\omega_2\frac{\partial  }{\partial \omega_2 })\\
		&\times(\frac{1 }{\Gamma(1-\alpha) }\int_1^\infty (s-1)^{-\alpha }s^{2\alpha-2}f_{\omega_1}(\omega_1 s^{\frac{\alpha }{3 }},\omega_2 s^{\frac{2\alpha }{3 }})\mathrm{d}s)\\
		=&(1-2\alpha-\frac{\alpha }{3}\omega_1\frac{\partial }{\partial \omega_1 }-\frac{2\alpha }{3}\omega_2\frac{\partial  }{\partial \omega_2 })\\
		&\times(\sum_{n,m=0}^{\infty}\frac{a_{n+1,m} \omega_1^n\omega_2^m }{\Gamma(1-\alpha) }\int_1^\infty(s-1)^{-\alpha}s^{\frac{(n+2m+6)\alpha }{3}-2}\mathrm{d}s) \\
		=&(1-2\alpha-\frac{\alpha }{3}\omega_1\frac{\partial }{\partial \omega_1 }-\frac{2\alpha }{3}\omega_2\frac{\partial  }{\partial \omega_2 })\\
		&\times(\sum_{n,m=0}^{\infty}\frac{a_{n+1,m} \omega_1^n\omega_2^m }{\Gamma(1-\alpha) }B(1-\alpha-\frac{n\alpha }{3 }-\frac{2m\alpha }{3 },1-\alpha)) \\
		=&(1-2\alpha-\frac{\alpha }{3}\omega_1\frac{\partial }{\partial \omega_1 }-\frac{2\alpha }{3}\omega_2\frac{\partial  }{\partial \omega_2 })\\
		&\times(\sum_{n,m=0}^{\infty}\frac{\Gamma(1-\alpha-\frac{n\alpha }{3 }-\frac{2m\alpha }{3 })}{\Gamma(2-2\alpha-\frac{n\alpha }{3 }-\frac{2m\alpha }{3 })}a_{n+1,m} \omega_1^n\omega_2^m) \\
		=&\sum_{n,m=0}^{\infty}\frac{\Gamma(1-\frac{(n+2m+3)\alpha }{3})}{\Gamma(1-\frac{(n+2m+6)\alpha }{3})}a_{n+1,m} \omega_1^n\omega_2^m, 
	\end{split} \tag{3.8}
\end{equation}
where $B(p,q)=\int_{0}^{1}x^{p-1}(1-x)^{q-1}dx$ is a beta function, and $\alpha$ must satisfy $1-\alpha-\frac{n\alpha }{3 }-\frac{2m\alpha }{3 }>0$ and $1-\alpha>0$. For the given $0<\alpha<1$, there exists two positive integers $n, m$ satisfy the condition $r: n + 2m < 3(1-\alpha)/\alpha$. If $\bar{r}: n + 2m \geqslant 3(1-\alpha)/\alpha$ holds, then the integral in (3.8) vanishes in the sense of the Hadamard's finite-part integral \cite{authour55}. Hadamard ignored the singular term and defined the integral as \cite{authour55} 
$$\int_a^b\frac{f(t) }{(b-t)^{\alpha+1}}\mathrm{d}t=\int_a^b\frac{f(t)-f(b) }{(b-t)^{\alpha+1}}\mathrm{d}t-\frac{f(b) }{\alpha(b-t)^{\alpha+1} },$$
where $0<\alpha<1$ and $f(t)$ is Lipschitz continuous on $[a,b]$. Coincidentally, the integrals in the sense of the Hadamard’s finite-part integral has the same results as the case of convergence \cite{authour52,authour53,authour54}.

Substituting (3.6)--(3.8) into the reduced equation (3.3) and equating the coefficients of different powers of $\omega$, we can obtain the explicit expressions of $a_{n,m}$ and $b_{n,m}$.
\begin{equation}
	\begin{split}
		&\sum_{n,m=0}^{r}\frac{\Gamma(1-\frac{(n+2m+3)\alpha }{3})}{\Gamma(1-\frac{(n+2m+6)\alpha }{3})}a_{n+1,m} \omega_1^n\omega_2^m-\sum_{n,m=0}^{\infty}\sum_{k=0}^{n}\sum_{j=0}^{m}(n-k+1)(k+1)a_{k+1,j}a_{n-k+1,m-j}\omega_1^n\omega_2^m\\
		&-\sum_{n,m=0}^{\infty}(n+4)(n+3)(n+2)(n+1)a_{n+4,m}\omega_1^n\omega_2^m-\sum_{n,m=0}^{\infty}(m+2)(m+1)a_{n,m+2}\omega_1^n\omega_2^m=0.
	\end{split} \tag{3.9}
\end{equation}
For $n + 2m < 3(1-\alpha)/\alpha$, we have
\begin{equation}
	\begin{split}
		a_{n+4,m}=&\frac{1 }{(n+4)(n+3)(n+2)(n+1) }\Big[\frac{\Gamma(1-\frac{(n+2m+3)\alpha }{3})}{\Gamma(1-\frac{(n+2m+6)\alpha }{3})}a_{n+1,m}\\
		&-(m+2)(m+1)a_{n,m+2}-\sum_{k=0}^{n}\sum_{j=0}^{m}(n-k+1)(k+1)a_{k+1,j}a_{n-k+1,m-j}\Big], 
	\end{split} \tag{3.10a}
\end{equation}
for $n + 2m \geqslant 3(1-\alpha)/\alpha$, we have
\begin{equation}
	\begin{split}
		a_{n+4,m}=&\frac{1 }{(n+4)(n+3)(n+2)(n+1) }\Big[-(m+2)(m+1)a_{n,m+2}\ \ \ \ \ \ \ \ \ \ \ \ \ \ \ \ \ \\
		&-\sum_{k=0}^{n}\sum_{j=0}^{m}(n-k+1)(k+1)a_{k+1,j}a_{n-k+1,m-j}\Big], 
	\end{split} \tag{3.10b}
\end{equation}
where $a_{n,m}=\frac{\partial^n }{\partial \omega_1^n}\frac{\partial^m }{\partial \omega_2^m}f(0,0)\ (n=0,1,2,3; m=0,1,2,\dots)$ are arbitrary constants. It means that the power series solution of (3.3) is
\begin{equation}
	\begin{split}
		f(\omega_1,\omega_2)=&\sum_{n=0}^{3}\sum_{m=0}^{\infty}a_{n,m} \omega_1^n\omega_2^m+\sum_{n,m=0}^{r}\frac{\omega^{n+4}_1\omega^{m}_2 }{(n+4)(n+3)(n+2)(n+1) }\\
		&\times\frac{\Gamma(1-\frac{(n+2m+3)\alpha }{3})}{\Gamma(1-\frac{(n+2m+6)\alpha }{3})}a_{n+1,m}-\sum_{n,m=0}^{\infty}\frac{\omega^{n+4}_1\omega^{m}_2 }{(n+4)(n+3)(n+2)(n+1) }\\
		&\times\Big[\sum_{k=0}^{n}\sum_{j=0}^{m}(n-k+1)(k+1)a_{k+1,j}a_{n-k+1,m-j}+(m+2)(m+1)a_{n,m+2}\Big]. 
	\end{split} \tag{3.11}
\end{equation}  

Therefore, the power series solution of Eq. (1.2) is
\begin{equation}
	\begin{split}
		u(t,x,y)=&\sum_{n=0}^{3}\sum_{m=0}^{\infty}a_{n,m} x^ny^mt^\frac{-(n+2m+2)\alpha }{3}+\sum_{n,m=0}^{r}\frac{x^{n+4}y^mt^\frac{-(n+2m+6)\alpha }{3} }{(n+4)(n+3)(n+2)(n+1) }\\
		&\times\frac{\Gamma(1-\frac{(n+2m+3)\alpha }{3})}{\Gamma(1-\frac{(n+2m+6)\alpha }{3})}a_{n+1,m}-\sum_{n,m=0}^{\infty}\frac{x^{n+4}y^mt^\frac{-(n+2m+6)\alpha }{3} }{(n+4)(n+3)(n+2)(n+1) }\\
		&\times\Big[\sum_{k=0}^{n}\sum_{j=0}^{m}(n-k+1)(k+1)a_{k+1,j}a_{n-k+1,m-j}+(m+2)(m+1)a_{n,m+2}\Big].
	\end{split} \tag{3.12}
\end{equation}

\noindent{\bf Theorem 3.2. }The power series solutions (3.12) are convergent in a neighborhood of the point $(0,0,|a_{0,0}|)$.

\bigskip
\noindent{\bf Proof: } From (3.10), we can obtain
\begin{equation}
	\begin{split}
		|a_{n+4,m}|\leqslant&\frac{1 }{(n+4)(n+3)(n+2)(n+1) }\Big[\Delta|a_{n+1,m}|+(m+2)(m+1)|a_{n,m+2}|\\
		&+\sum_{k=0}^{n}\sum_{j=0}^{m}(n-k+1)(k+1)|a_{k+1,j}||a_{n-k+1,m-j}|\Big], 
	\end{split} \tag{3.13}
\end{equation}
where $\Delta=\frac{|\Gamma(1-\frac{(n+2m+3)\alpha }{3})|}{|\Gamma(1-\frac{(n+2m+6)\alpha }{3})|}$ for $n + 2m < 3(1-\alpha)/\alpha$, $\Delta=0$ for $n + 2m \geqslant 3(1-\alpha)/\alpha$. Thus, for arbitrary natural numbers $n$ and $m$, (3.13) can be written as
\begin{equation}
	|a_{n+4,m}|\leqslant M\Big(|a_{n+1,m}|+\sum_{k=0}^{n}\sum_{j=0}^{m}|a_{k+1,j}||a_{n-k+1,m-j}|+|a_{n,m+2}|\Big),  \tag{3.14}
\end{equation}
where $M=\mathrm{max}\{\frac{|\Gamma(1-\frac{(n+2m+3)\alpha }{3})|}{(n+4)(n+3)(n+2)(n+1)|\Gamma(1-\frac{(n+2m+6)\alpha }{3})|},\frac{1 }{(n+4)(n+3) }\}$.

Consider another power series
$$P(\omega_1,\omega_2)=\sum_{n,m=0}^{\infty}p_{n,m} \omega_1^n\omega_2^m,\eqno{(3.15)}$$ 
where $p_{n,m}=|a_{n,m}|\ (n=0,1,2,3;m=0,1,2,\dots)$ and 
\begin{equation}
	p_{n+4,m}=M\Big(p_{n+1,m}+\sum_{k=0}^{n}\sum_{j=0}^{m}p_{k+1,j}p_{n-k+1,m-j}+p_{n,m+2}\Big).  \tag{3.16}
\end{equation}
Therefore, it is easily seen that $|a_{n,m}|\leq p_{n,m}$ and $|b_{n,m}|\leq q_{n,m}$ for $n,m=0,1,2,\ldots$, that is, the power series (3.15) is the majorant series of (3.6). We next show that the power series (3.15) is convergent. By simple calculation, we can
get
\begin{equation}
	\begin{split}
		P(\omega_1,\omega_2)=&\sum_{n=0}^{3}\sum_{m=0}^{\infty}p_{n,m} \omega_1^n\omega_2^m+M\big((P(\omega_1,\omega_2)-\sum_{m=0}^{\infty}p_{0,m}\omega_2^m)\omega_1^3\\
		&+(P(\omega_1,\omega_2)-\sum_{m=0}^{\infty}p_{0,m}\omega_2^m)(P(\omega_1,\omega_2)-\sum_{m=0}^{\infty}p_{0,m}\omega_2^m)\omega_1^2\\
		&+(P(\omega_1,\omega_2)-\sum_{n=0}^{\infty}\sum_{m=0}^{1}p_{n,m}\omega_1^n\omega_2^m)\omega_1^4\omega_2^{-2}\big). 
	\end{split} \tag{3.17}
\end{equation}
Consider the following implicit function with respect to the independent variable $\omega_1, \omega_2$:
\begin{equation}
	\begin{split}
		F(\omega_1,\omega_2,P)=&P-\sum_{n=0}^{3}\sum_{m=0}^{\infty}p_{n,m} \omega_1^n\omega_2^m-M\big((P-\sum_{m=0}^{\infty}p_{0,m}\omega_2^m)\omega_1^3\\
		&+(P-\sum_{m=0}^{\infty}p_{0,m}\omega_2^m)(P-\sum_{m=0}^{\infty}p_{0,m}\omega_2^m)\omega_1^2\\
		&+(P-\sum_{n=0}^{\infty}\sum_{m=0}^{1}p_{n,m}\omega_1^n\omega_2^m)\omega_1^4\omega_2^{-2}\big),
	\end{split} \tag{3.18}
\end{equation}
which is analytic in a neighborhood of $(0, 0, p_{0,0})$, and $F(0, 0, p_{0,0})=0$,  $\frac{\partial F }{\partial P }|_{(0, 0, p_{0,0})}=1\neq0$. Therefore, by implicit function theorem, the power series (3.15) is analytic in neighborhood of the point $(0, 0, p_{0,0})$. It implies that the power series solution (3.6) is convergent in a neighborhood of the point $(0, 0, |a_{0,0}|)$. This completes the proof. \hfill{$\square$}

\bigskip
\noindent{\bf Case 2: $X_2=\frac{\partial }{\partial x }$}
\bigskip

The characteristic equation corresponding to the group generator $X_2$ is
$$\frac{\mathrm{d} t }{0}=\frac{\mathrm{d}x }{1 }=\frac{\mathrm{d} y }{0 }=\frac{\mathrm{d}u }{0 },\eqno{(3.19)}$$
from which, we obtain the similarity variables $t$, $y$ and $u$. So we get the invariant solution of Eq. (1.2) as follows:
$$u=f(t,y).\eqno{(3.20)}$$
Substituting (3.20) into Eq. (1.2), we have the following reduced equation:
$$f_{yy}=0,\eqno{(3.21)} $$
from which, we can obtain the following trivial solution:
$$u=g(t)y+h(t),\eqno{(3.22)}$$
where $g(t)$ and $h(t)$ are arbitrary functions.

\bigskip
\noindent{\bf Case 3: $X_3=\frac{\partial }{\partial y }$}
\bigskip

The characteristic equation corresponding to the group generator $X_3$ is
$$\frac{\mathrm{d} t }{0}=\frac{\mathrm{d}x }{0 }=\frac{\mathrm{d} y }{1 }=\frac{\mathrm{d}u }{0 },\eqno{(3.23)}$$
from which, we obtain the similarity variables $t$, $x$ and $u$. So we get the invariant solution of Eq. (1.2)
$$u=u(t,x),\eqno{(3.24)}$$
and the reduced equation
$$D^{\alpha}_{t}u_{x}-uu_{xx}-u^2_{x}-u_{xxxx}=0,\eqno{(3.25)} $$
which is (1+1)-dimensional time fractional partial differential equation. For (3.25), we can once again use the Lie symmetry analysis method to obtain the following generators: 
$$\varLambda_1=t\frac{\partial }{\partial t }+\frac{\alpha }{3} x\frac{\partial }{\partial x }-\frac{2\alpha }{3}u\frac{\partial }{\partial u },\ \ \varLambda_2=\frac{\partial }{\partial x }.\eqno{(3.26)} $$

(1) For $\varLambda_1$, the characteristic equation is
$$\frac{\mathrm{d} t }{t}=\frac{\mathrm{d}x }{\frac{\alpha }{3} x }=\frac{\mathrm{d}u }{-\frac{2\alpha }{3}u },\eqno{(3.27)}$$
from which, we obtain the similarity variables $xt^{-\frac{\alpha }{3}}$ and $ut^\frac{2\alpha }{3}$. So we have the following invarant solutions:
$$u=t^{-\frac{2\alpha }{3}}f(\omega),\ \ \omega=xt^{-\frac{\alpha }{3}}.\eqno{(3.28)}$$
Substituting (3.28) into Eq. (3.25), we can get the following equation:
$$(\mathcal{P}^{1-2\alpha,\alpha}_{\frac{3 }{\alpha}}f')(\omega)-ff''-(f')^2-f^{(4)}=0.\eqno{(3.29)} $$

Next we use the power series method to derive the power series solutions of the reduced equations (3.21). Let us assume
$$f(\omega)=\sum_{k=0}^{\infty}a_k \omega^k,\eqno{(3.30)}$$ 
then
\begin{equation}
	\begin{split}
		&f'(\omega)=\sum_{k=0}^{\infty}(k+1)a_{k+1} \omega^k,\\ &f''(\omega)=\sum_{k=0}^{\infty}(k+2)(k+1)a_{k+2} \omega^k, \\
		&f^{(4)}(\omega)=\sum_{k=0}^{\infty}(k+4)(k+3)(k+2)(k+1)a_{k+4} \omega^k,
	\end{split} \tag{3.31}
\end{equation}
and
\begin{equation}
	(\mathcal{P}^{1-2\alpha,\alpha}_{\frac{3 }{\alpha}}f')(\omega)=\sum_{k=0}^{\infty}\frac{\Gamma(1-\frac{(k+3)\alpha }{3 })}{\Gamma(1-\frac{(k+6)\alpha }{3 })}(k+1)a_{k+1} \omega^k, 
	\tag{3.32}
\end{equation}
with the conditions $1-\frac{(k+3)\alpha }{3 }>0$ and $1-\alpha>0$. For the given $0<\alpha<1$, there exists a positive integer $s$ such that for all integers $k\leqslant s$, $(k+3)\alpha < 3$ and the integral in (3.32) exists. If $k> s$, then $(k+3)\alpha \geqslant 3$ and the integral in (3.32) vanishes in the sense of the Hadamard's finite-part integral. 

Substituting (3.30)--(3.32) into the reduced equation (3.25) and equating the coefficients of different powers of $\omega$, we can obtain the explicit expressions of $a_k$. 
For $k\leqslant s$, we have
\begin{equation}
	\begin{split}
		a_{k+4}=&\frac{1 }{(k+4)(k+3)(k+2)(k+1) }\Big[\frac{\Gamma(1-\frac{(k+3)\alpha }{3 })}{\Gamma(1-\frac{(k+6)\alpha }{3 })}(k+1)a_{k+1}\\
		&-\sum_{i+j=k}^{}(i+2)(i+1)a_{i+2}a_{j}-\sum_{i+j=k}^{}(i+1)(j+1)a_{i+1}a_{j+1}\Big], 
	\end{split} \tag{3.33a}
\end{equation}
for $k>s$, we have
\begin{equation}
	\begin{split}
		a_{k+4}=&\frac{1 }{(k+4)(k+3)(k+2)(k+1) }\Big[-\sum_{i+j=k}^{}(i+2)(i+1)a_{i+2}a_{j}\\
		&-\sum_{i+j=k}^{}(i+1)(j+1)a_{i+1}a_{j+1}\Big], 
	\end{split} \tag{3.33b}
\end{equation}
where $a_i=f^{(i)}(0),(i=0,1,2,3)$ are arbitrary constants. It means that the power series solution of (3.25) is
\begin{equation}
	\begin{split}
		f(\omega)=&a_0+a_1\omega+a_2\omega^2+a_3\omega^3+\sum_{k=0}^{s}\frac{\omega^{k+4} }{(k+4)(k+3)(k+2)(k+1)}\\
		&\times\frac{\Gamma(1-\frac{(k+3)\alpha }{3 })}{\Gamma(1-\frac{(k+6)\alpha }{3 })}(k+1)a_{k+1}-\sum_{k=0}^{\infty}\frac{\omega^{k+4} }{(k+4)(k+3)(k+2)(k+1)}\\
		&\times\Big[\sum_{i+j=k}^{}(i+2)(i+1)a_{i+2}a_{j}+\sum_{i+j=k}^{}(i+1)(j+1)a_{i+1}a_{j+1}\Big].
	\end{split} \tag{3.34}
\end{equation}  

Therefore, the power series solution of Eq. (1.2) is
\begin{equation}
	\begin{split}
		u(t,x,y)=&a_0t^{-\frac{2\alpha}{3}}+a_1xt^{-\alpha}+a_2x^2t^{-\frac{4\alpha}{3}}+a_3x^3t^{-\frac{5\alpha}{3}}+\sum_{k=0}^{s}\frac{x^{k+4}t^{-\frac{(k+6)\alpha}{3}} }{(k+4)(k+3)(k+2)(k+1)} \\
		&\times\frac{\Gamma(1-\frac{(k+3)\alpha }{3 })}{\Gamma(1-\frac{(k+6)\alpha }{3 })}(k+1)a_{k+1}-\sum_{k=0}^{\infty}\frac{x^{k+4}t^{-\frac{(k+6)\alpha}{3}} }{(k+4)(k+3)(k+2)(k+1)} \\
		&\times\Big[\sum_{i+j=k}^{}(i+2)(i+1)a_{i+2}a_{j}+\sum_{i+j=k}^{}(i+1)(j+1)a_{i+1}a_{j+1}\Big].
	\end{split} \tag{3.35}
\end{equation} 

\noindent{\bf Theorem 3.3. }The power series solutions (3.35) are convergent in a neighborhood of the point $(0, |a_0|)$.

\bigskip
\noindent{\bf Proof: } From (3.33), we can obtain
\begin{equation}
	\begin{split}
		|a_{k+4}|\leq&\frac{1 }{(k+4)(k+3)(k+2)(k+1) }\Big[\Delta(k+1)|a_{k+1}|\\
		&+\sum_{i+j=k}^{}(i+2)(i+1)|a_{i+2}||a_{j}|+\sum_{i+j=k}^{}(i+1)(j+1)|a_{i+1}||a_{j+1}|\Big], 
	\end{split} \tag{3.36}
\end{equation}
where $\Delta=\frac{|\Gamma(1-\frac{(k+3)\alpha }{3 })|}{|\Gamma(1-\frac{(k+6)\alpha }{3 })|}$ for $k\leqslant s$, $\Delta=0$ for $k> s$. Thus, for arbitrary natural numbers $k$, (3.36) can be written as
\begin{equation}
	|a_{k+4}|\leq M\Big[|a_{k+1}|+\sum_{i+j=k}^{}|a_{i+2}||a_{j}|+\sum_{i+j=k}^{}|a_{i+1}||a_{j+1}|\Big], 
	\tag{3.37}
\end{equation}
where $M=\mathrm{max}\{\frac{|\Gamma(1-\frac{(k+3)\alpha }{3 })|}{(k+4)(k+3)(k+2)|\Gamma(1-\frac{(k+6)\alpha }{3 })|},\frac{1 }{(k+4)(k+3) }\}$.

Consider another power series
$$B(\omega)=\sum_{k=0}^{\infty}b_k \omega^k,\eqno{(3.38)}$$
where $b_i=|a_{i}|\  (i=0,1,2,3)$ and
$$b_{k+4}= M(b_k+\sum_{i+j=k}^{}b_{i+2}b_{j}+\sum_{i+j=k}^{}b_{i+1}b_{j+1}),\ \ k\geq0.\eqno{(3.39)}$$
Therefore, it is easily seen that $|a_{k}|\leq b_{k}$ for $k=0,1,2,\ldots$, that is, the power series (3.38) is the majorant series of (3.30). We next show that the power series (3.38) is convergent. By simple calculation, we can
get
\begin{equation}
	B(\omega)=b_0+b_1\omega+b_2\omega^2+b_3\omega^3+M\omega^2\big(B(\omega)\omega^2+B(\omega)(B(\omega)-b_0-b_1\omega)+(B(\omega)-b_0)^2\big).
	\tag{3.40}
\end{equation}
Consider the following implicit function with respect to the independent variable $\omega$:
\begin{equation}
	\Psi(\omega,B)=B-b_0-b_1\omega-b_2\omega^2-b_3\omega^3-M\omega^2\big(B\omega^2+B(B-b_0-b_1\omega)+(B-b_0)^2\big). \tag{3.41}
\end{equation}
which is analytic in a neighborhood of $(0, b_0)$, and $\Psi(0,b_0)=0$, $\frac{\partial }{\partial B }\Psi(0,b_0)=1$. Therefore, by implicit function theorem, the power series (3.38) is analytic in neighborhood of the point $(0, b_0)$. It implies that the power series solution (3.35) is convergent in a neighborhood of the point $(0, |a_0|)$. This completes the proof. \hfill{$\square$}

(2) For $\varLambda_2$, the characteristic equation is
$$\frac{\mathrm{d} t }{0}=\frac{\mathrm{d}x }{x }=\frac{\mathrm{d}u }{0 },\eqno{(3.42)}$$
from which, we obtain the similarity variables $t$ and $u$. So we have the following invarant solution:
$$u=f(t),\eqno{(3.43)}$$
which satisfies Eq. (1.2), and is one of its trivial solutions.

\bigskip
\noindent{\bf Case 4: $X_2+X_3=\frac{\partial }{\partial x }+\frac{\partial }{\partial y }$}
\bigskip

The characteristic equation corresponding to the group generator $X_2+X_3$ is
$$\frac{\mathrm{d} t }{0}=\frac{\mathrm{d}x }{1 }=\frac{\mathrm{d} y }{1 }=\frac{\mathrm{d}u }{0 },\eqno{(3.44)}$$
from which, we obtain the similarity variables $t$, $x-y$ and $u$. So we get the invariant solution of Eq. (1.2)
$$u=u(t,\omega),\ \ \omega=x-y,\eqno{(3.45)}$$
and the reduced equation
$$D^{\alpha}_{t}u_{\omega}-uu_{\omega\omega}-u^2_{\omega}-u_{\omega\omega\omega\omega}+u_{\omega\omega}=0,\eqno{(3.46)} $$
which is (1+1)-dimensional time fractional partial differential equation. For (3.46), we use Lie symmetry analysis method once again to obtain the following generator: 
$$\varLambda_1=\frac{\partial }{\partial \omega }.\eqno{(3.47)} $$
However, in this case, we only obtain one trivial solution as (3.43).

 \section{Conservation laws of Eq. (1.2)} 

In this section, we will construct conservation laws of Eq. (1.2) by using the generalization of the Noether operators and the new conservation theorem.

The Eq. (1.2) is denoted as
$$F=D^{\alpha}_{t}u_{x}-uu_{xx}-u^2_{x}-u_{xxxx}-u_{yy}=0,\eqno{(4.1)} $$
of which the formal Lagrangian \cite{authour56} is given by
\begin{equation}
	\mathcal{L}=p(t,x,y)F=p(t,x,y)\big(D^{\alpha}_{t}u_{x}-uu_{xx}-u^2_{x}-u_{xxxx}-u_{yy}\big), 
	\tag{4.2}
\end{equation}
where $p(t,x,y)$ is a new dependent variable. The Euler-Lagrange operator is
$$\frac{\delta}{\delta u}=\frac{\partial }{\partial u }-(D^{\alpha}_{t})^*D_{x}\frac{\partial }{\partial (D^{\alpha}_{t}u_x) }-D_{x}\frac{\partial }{\partial u_{x} }+D^2_{x}\frac{\partial }{\partial u_{xx} }+D^2_{y}\frac{\partial }{\partial u_{yy} }+D^4_{x}\frac{\partial }{\partial u_{xxxx} },\eqno{(4.3)} $$
where $(D^{\alpha}_{t})^*$ is the adjoint operator of $D^{\alpha}_{t}$. It is defined by the right-sided of Caputo fractional derivative, i.e., \cite{authour57}
$$(D^{\alpha}_{t})^*f(t,x)\equiv\ _{t}^{c}D_{T}^{\alpha}f(t,x)= \left\{
\begin{array}{lll}
	\frac{1}{\Gamma (n-\alpha )}\int_t^T {\frac{1}{(t-s)^{ \alpha-n+1 }} }\frac{\partial^n }{\partial {s^n}}f(s,x)\mathrm{d}s,& n-1<\alpha<n, n\in\mathbb{N} \\
	D_{t}^{n}f(t,x),& \alpha=n\in\mathbb{N}.  \\
\end{array}\right. \eqno{(4.4)}$$
The adjoint equation to (4.1) is given by
$$F^*=\frac{\delta \mathcal{L} }{\delta u}=-(D^{\alpha}_{t})^*p_{x}-p_{xxxx}-p_{yy}-up_{xx}-3p_{x}u_{x}=0,\eqno{(4.5)} $$
from which, we can obtain the non-zero solution $p=\varphi(t)y$ with an arbitrary differentiable function $\varphi(t)$, confirming the nonlinear self-adjointness of Eq. (1.2).

Next we will use the above adjoint equation and the new conservation theorem to construct conservation laws of Eq. (1.2). From the classical definition of the conservation laws, a vector $C=(C^t,C^x,C^y)$ is called the conserved vector for the governing equation if it satisfies the conservation equation $[D_tC^t+D_xC^x+D_yC^y]_{F=0}=0$. We can obtain the components of the conserved vector by using the generalization of the Noether operators.

Firstly, from the fundamental operator identity, i.e., (see \cite{authour57,authour58})
$$prX+D_t\tau\cdot\mathcal{I}+D_x\xi\cdot\mathcal{I}+D_y\theta\cdot\mathcal{I}=W^u\cdot\frac{\delta}{\delta u}+D_t\mathcal{N}^t+D_x\mathcal{N}^x+D_y\mathcal{N}^y,\eqno{(4.6)}  $$
where $prX$ is mentioned in (2.3), $\mathcal{I}$ is the identity operator and $W=\eta-\tau u_t-\xi u_x-\theta u_y$ is the characteristic for group generator $X$, we can get fractional Noether operators \cite{authour59} as follows:
$$\mathcal{N}^t=\tau\mathcal{I}-\sum_{k=0}^{n-1 }(-1)^kD_{t}^{\alpha-1-k}(W)D_{t}^kD_{x}\frac{\partial }{\partial (D^{\alpha}_{t}u_x) }+(-1)^nJ(W,D_t^nD_{x}\frac{\partial }{\partial (D^{\alpha}_{t}u_{x}) }), \eqno{(4.7)}  $$
\begin{equation}
	\begin{split}
		\mathcal{N}^x=&\xi\mathcal{I}+D_{t}^{\alpha}(W)\frac{\partial }{\partial (D^{\alpha}_{t}u_x) }+W\big(\frac{\partial }{\partial u_x }-D_{x}\frac{\partial }{\partial u_{xx}}-D^3_{x}\frac{\partial }{\partial u_{xxxx}}\big)\\
		&+D_{x}(W)\big(\frac{\partial }{\partial u_{xx} }+D^2_{x}\frac{\partial }{\partial u_{xxxx}}\big)-D^2_{x}(W)D_{x}\frac{\partial }{\partial u_{xxxx} }+D^3_{x}(W)\frac{\partial }{\partial u_{xxxx} }, \ \ \ \ \ \ \ \ \end{split} \tag{4.8}
\end{equation}
$$\mathcal{N}^y=\theta\mathcal{I}-WD_{y}\frac{\partial }{\partial u_{yy} }+D_{y}(W)\frac{\partial }{\partial u_{yy}}, \ \ \ \ \ \ \ \ \ \ \ \ \ \ \ \ \ \ \ \ \ \ \ \ \ \ \ \ \ \ \ \ \ \ \ \ \ \ \ \  \ \ \ \ \ \ \ \ \ \ \ \ \ \ \ \eqno{(4.9)}  $$
where $n=[\alpha]+1$ and $J$ is given by
$$J(f,g)=\frac{1 }{\Gamma(n-\alpha) }\int_0^t\int_t^T\frac{f(\tau,x,y)g(\theta,x,y) }{(\theta-\tau)^{\alpha+1-n} }\mathrm{d}\theta\mathrm{d}\tau. \eqno{(4.10)}  $$
The components of conserved vector are defined by \cite{authour59}
$$C^t=\mathcal{N}^t\mathcal{L},\ \ C^x=\mathcal{N}^x\mathcal{L},\ \ C^y=\mathcal{N}^y\mathcal{L}. \eqno{(4.11)}  $$

\noindent{\bf Case 1: $X_1=t\frac{\partial }{\partial t }+\frac{\alpha }{3 }x\frac{\partial }{\partial x }+\frac{2\alpha }{3 } y\frac{\partial }{\partial y }-\frac{2\alpha }{3 }u\frac{\partial }{\partial u }$}
\bigskip

The characteristics of $X_1$ are
$$W=-\frac{2\alpha }{3 }u-tu_t-\frac{\alpha }{3 }xu_x-\frac{2\alpha }{3 } yu_y. \eqno{(4.12)}$$
Therefore, for $0<\alpha<1$,
\begin{equation}
	\begin{split}
		C^t=&0,\\
		C^x=&-\varphi(t)yD_{t}^{\alpha}(\frac{2\alpha }{3 }u+tu_t+\frac{\alpha }{3 }xu_x+\frac{2\alpha }{3 } yu_y)+\varphi(t)yu_x(\frac{2\alpha }{3 }u+tu_t\\
		&+\frac{\alpha }{3 }xu_x+\frac{2\alpha }{3 } yu_y)+\varphi(t)yu(\alpha u_x+tu_{xt}+\frac{\alpha }{3 }xu_{xx}+\frac{2\alpha }{3 } yu_{xy})\\
		&+\varphi(t)y(\frac{5\alpha }{3 }u_{xx}+tu_{xxxt}+\frac{\alpha }{3 }xu_{xxxx}+\frac{2\alpha }{3 } yu_{xxxy}), \\
		C^y=&-\varphi(t)(\frac{2\alpha }{3 }u+tu_t+\frac{\alpha }{3 }xu_x+\frac{2\alpha }{3 } yu_y)\\
		&+\varphi(t)y(\frac{4\alpha }{3 }u_y+tu_{ty}+\frac{\alpha }{3 }xu_{xy}+\frac{2\alpha }{3 } yu_{yy}).
	\end{split}\tag{4.13}
\end{equation}

\noindent{\bf Case 2: $X_2=\frac{\partial }{\partial x }$}
\bigskip

The characteristics of $X_2$ are
$$W=-u_x. \eqno{(4.14)}$$
Therefore, for $0<\alpha<1$,
\begin{equation}
	\begin{split}
		C^t=&0,\\
		C^x=&\varphi(t)y(-D_{t}^{\alpha}u_x+u_x^2+uu_{xx}+u_{xxxx}), \\
		C^y=&-\varphi(t)u_x+\varphi(t)yu_{xy}.
	\end{split}\tag{4.15}
\end{equation}

\noindent{\bf Case 3: $X_3=\frac{\partial }{\partial y }$}
\bigskip

The characteristics of $X_3$ are
$$W=-u_y. \eqno{(4.16)}$$
Therefore, for $0<\alpha<1$,
\begin{equation}
	\begin{split}
		C^t=&0,\\
		C^x=&\varphi(t)y(-D_{t}^{\alpha}u_y+u_xu_y+uu_{xy}+u_{xxxy}), \\
		C^y=&-\varphi(t)u_y+\varphi(t)yu_{yy}.
	\end{split}\tag{4.17}
\end{equation}
 
\section{Conclusion} 
 
This paper demonstrates that Lie symmetry analysis method is effective for studying nonlinear partial differential equations with the mixed derivative of Riemann-Liouville time-fractional derivative and integer-order $x$-derivative. As an example, this article studies the (2+1)-dimensional time fractional Kadomtsev-Petviashvili equation, obtains its Lie symmetries, and repeatedly uses the Lie symmetry analysis method to reduce the KP equation and obtain some exact solutions. From the convergent power series solutions (3.12) and (3.35), we can see that the fractional order plays a crucial role in the obtained results. To our knowledge, the order $\alpha$ of the fractional derivative in fractional diffusion-type equations determines the types of diffusion, including subdiffusion ($\alpha\in(0,1)$), normal diffusion ($\alpha=1$) and wave diffusion ($\alpha\in(1,2)$) \cite{authour41}. Therefore, fractional derivative can more accurately describe anomalous diffusion and wave propagation in different media. It means that the results obtained in this paper can more accurately reflect the dynamic behavior of the waves described by the studied equation based on the choice of order $\alpha$. As the Lie symmetry analysis method is widely applied to time-fractional or space-fractional differential equations, it is also gradually applied to FDEs with the mixed derivative of time-fractional derivative and integer-order $x$-derivative. Inspired by these research advances, our next research goal is to use the Lie symmetry analysis method to investigate FDEs with the mixed derivative of time-fractional derivative and space-fractional derivative ($\frac{\partial^\alpha}{\partial t^\alpha}\frac{\partial^\beta u}{\partial x^\beta}$).

\subsection*{Acknowledgements}

The authors would like to thank the anonymous reviewers for their careful reading of
the previous version of the paper and for their comments and suggestions which have improved the paper very much.

\label{lastpage}
\end{document}